\documentclass[%
 aip,
 amsmath,amssymb,
 reprint,%
]{revtex4-1}

\usepackage[utf8]{inputenc}
\usepackage[T1]{fontenc}
\usepackage{bm}

\usepackage{csquotes}

\usepackage{graphicx}
\usepackage{overpic}

\usepackage{amsmath}
\usepackage{amssymb}
\usepackage{amsthm}
\usepackage{diffcoeff}
\usepackage{dsfont}

\usepackage[hidelinks]{hyperref}
\usepackage[nameinlink]{cleveref}

\usepackage[normalem]{ulem}

\usepackage{xcolor}

\renewcommand{\u}{\mathbf{u}}
\renewcommand{\v}{\mathbf{v}}
\newcommand{\w}{\mathbf{w}}

\newcommand{\xixi}{\bm{\xi}}

\newcommand*{\R}{\ensuremath{\mathbb{R}}}
\newcommand*{\C}{\ensuremath{\mathbb{C}}}

\NewDocumentCommand{\mathfunc}{mo}{\ensuremath{\mathop{#1{}} \IfValueT{#2}{\mathopen{} \left( #2 \right)}}}
\newcommand{\fn}{\mathfunc}
\newcommand{\definemathfunc}[2]{
  \NewDocumentCommand{#1}{o}{%
    \ensuremath{%
      \mathop{#2{}} \IfValueT{##1}{\mathopen{} \left( ##1 \right)}
    }
  }
}
\NewDocumentCommand{\cont}{o}{\ensuremath{\mathcal{C}\IfValueT{#1}{^#1}}}
\renewcommand*{\vec}[1]{\ensuremath{\boldsymbol{#1}}}
\newcommand*{\onevec}{\ensuremath{\mathds{1}}}

\newcommand*{\der}[2]{\ensuremath{#1^{(#2)}}}
\begin{document}

% === Metadata ===
\title{Exact Dimensional Reduction for Quasi-Linear ODE Ensembles}
\date{\today}

\author{Felix Augustsson}
\affiliation{Centre for Mathematical Sciences, Lund University, M\"arkesbacken 4, 223 62 Lund, Sweden}
% \email{felix.augustsson@math.lth.se; rok.cestnik@math.lth.se; erik.martens@math.lth.se}

\author{Erik A. Martens}
\affiliation{Centre for Mathematical Sciences, Lund University, M\"arkesbacken 4, 223 62 Lund, Sweden}
\affiliation{IMFUFA, Department of Science and Environment, Roskilde University, Universitetsvej 1, Roskilde, Denmark}

\author{Rok Cestnik}
\affiliation{Centre for Mathematical Sciences, Lund University, M\"arkesbacken 4, 223 62 Lund, Sweden}

\begin{abstract}
We present an exact dimensional reduction for high-dimensional dynamical systems composed of $N$ identical dynamical units governed by quasi-linear ordinary differential equations (ODEs) of order $M$.
In these systems, each unit follows a linear differential equation whose coefficients depend nonlinearly on the ensemble variables, such as a mean field variable.
We derive $M+1$ closed-form macroscopic equations of order $M$ with variables that exactly capture the full microscopic dimensional dynamics and that allow reconstruction of individual trajectories from the reduced system. 
Our approach enables low-dimensional analysis of collective behavior in coupled oscillator networks and provides computationally efficient exact representations of large-scale dynamics.
We illustrate the theory with examples, highlighting new families of solvable models relevant to physics, biology and engineering that are now amenable to simplified analysis.
\end{abstract}

% === Start of paper ===
\maketitle

\begin{quotation}
 Understanding how to extract macroscopic dynamics from high-dimensional systems composed of interacting units remains a central challenge across nonlinear science.
 In the context of coupled oscillators, paradigmatic models such as the Kuramoto system,\cite{kuramoto1984,strogatz2000,pikovsky2003,arenas2008} have delivered deep insights into the emergence of synchronization processes and other collective behaviors.
 A key breakthrough in this area was the discovery of exact dimensional reductions, notably the Watanabe-Strogatz theory\cite{watanabe1994,strogatz2000} and the Ott-Antonsen ansatz,\cite{ott2008} which yield closed-form macroscopic equations that exactly determine the microscopic dynamics.
 These ideas have been further developed in works linking oscillator networks to Möbius group symmetries and ensembles of Riccati-like equations.\cite{marvel2009,chen2017,CestnikMartens2024}
 In our work, we extend these methods to another family of systems --- namely, quasi-linear ordinary differential equations --- in which linear derivatives are coupled through coefficients depending on a macroscopic mean field, a quantity given by knowledge of the microscopic variables.
 We provide a general method for exact reduction of quasi-linear systems and illustrate its application through analytically tractable examples.
\end{quotation}

\section{Introduction}

% Network Dynamics and Collective Behavior
The dynamics of large networks of interacting units are central to many fields, ranging from physics and biology to engineering and the social sciences.
A hallmark of such systems is the emergence of macroscopic order --- such as synchronization, clustering, or swarming --- arising from simple rules on the microscopic level.
Prominent examples include phase oscillators in neuroscience and physics,\cite{kuramoto1975,strogatz2000,pikovsky2001} coordination in animal groups,\cite{vicsek1995} power grid structures\cite{rohden2012} and epidemic dynamics on complex networks.\cite{arenas2008}
A major challenge in these settings is to elucidate how  large-scale collective behavior arises from the underlying microscopic dynamics.
In particular, for oscillator networks, the ability to derive reduced macroscopic descriptions that exactly capture the full system dynamics is of great interest.

% Dimensional Reduction in Oscillator Networks.
In coupled phase oscillator systems, exact dimensional reduction techniques have enabled substantial analytical progress.
Foundational approaches, such as the Watanabe--Strogatz theory and the Ott--Antonsen ansatz, provide exact, closed-form reductions of high-dimensional dynamics to low-dimensional manifolds under suitable symmetry assumptions.\cite{watanabe1994,ott2008,Bick2018c}
These reductions often proceed via a mean field framework, yielding macroscopic equations that not only describe the evolution of collective observables but also exactly determine the orbits of the underlying microscopic system. 
Subsequent work has deepened and generalized these ideas, providing dimensional reductions for ensembles of Riccati-like equations whose dynamics are governed by Möbius group symmetries structure~\cite{marvel2009,chen2017,CestnikMartens2024}. 
Together, these frameworks illustrate how exact reductions link microscopic and macroscopic descriptions, thereby facilitating a simplified analysis of oscillator networks based on low dimensional descriptions.

% Beyond Riccati—Our Contribution. Linkage Riccati to 2nd order and beyond.
In a recent study, dimensional reductions were extended to complex-valued Riccati-like systems,\cite{CestnikMartens2024} paving the way for the treatment of a broader class of dynamics in the complex plane.
That study also observed a classical connection between the Riccati equation and second-order linear differential equations via a variable transformation,\cite{inceordinary} and pointed out that a similar transformation may exist that maps ensembles of Riccati-like equations to quasi-linear ensembles with nonlinear mean field coupling (see App.~\ref{sec:riccati-linkage}).

In this work, taking inspiration from this observation, we develop a formalism for exact dimensional reduction for a class of systems, which we refer to as \emph{quasi-linear ordinary differential equations}. 
These systems consist of ensembles of \emph{identical oscillatory units} whose governing equations remain linear in their derivatives, while their coefficients are functions that depend on a macroscopic mean field quantity constructed from the microscopic variables.
This quasi-linear structure retains essential nonlinearity while preserving sufficient linearity to enable tractable reduction.
While exact dimensional reductions have been well studied for the Kuramoto and Theta neuron model, both of which are based on Riccati-like dynamics, they remain largely unexplored for the class of quasi-linear systems. 
Our framework identifies the conditions under which such reductions are possible and provides a methodology for deriving low-dimensional, closed-form equations that fully capture the high-dimensional dynamics.

% Overview
We first motivate the dimensional reduction procedure using a simple second-order quasi-linear system in Sec.~\ref{sec:motivating-example}.
Then we generalize this result to quasi-linear systems of arbitrary order in Sec.~\ref{sec:general-result}.
In Sec.~\ref{sec:special-global-dependence}, we explain how dependence on a mean-field coupling with weighted sums allows for a parameter-independent formulation of the macroscopic description. 
We demonstrate the applicability of our framework on two concrete examples in Sec.~\ref{sec:examples}, and conclude with a discussion in Sec.~\ref{sec:discussion}.
Finally, while the reduction method stands in its own right, we elucidate the connection between quasi-linear systems and Riccati-like equations that were explored in earlier works, see App.~\ref{sec:riccati-linkage}.

% ================================================================

\section{Dimensional reduction for second order quasi-linear systems}
\label{sec:motivating-example}
We consider an ensemble of \emph{second order} ordinary differential equations in $ \u(t) = (u_1(t), \dots, u_N(t))$, $\u:\R \to \C^N$, which we refer to as the \emph{microscopic equations},
\begin{equation}
  \ddot{u}_k + \fn{a_1}[\u, t] \dot{u}_k + \fn{a_0}[\u, t] u_k = \fn{g}[\u, t]
  \label{eq:microscopic_sec_order}
\end{equation}
for $ k = 1,\dots,N $, where the coefficients $a_0: \C^N\times\R\to \C , a_1:\C^N\times\R\to \C$ and the inhomogeneous term $g:\C^N\times\R\to \C$ are scalar complex-valued functions.
Observe that these functions are \emph{identical} for all nodes $ k $.
Note that system is composed of $N$ equations of order 2, amounting to a $2N$ solution space.

Derivatives in \eqref{eq:microscopic_sec_order} appear as linear terms, while nonlinearity may enter through the dependence of the coefficients $a_1,a_0$ and $g$ on $\u$.
We therefore refer to the structure of \eqref{eq:microscopic_sec_order} as a \emph{quasi-linear system}.
In analogy to linear systems, we refer to the system as homogeneous if $g\equiv 0$, and inhomogeneous otherwise.

A physical interpretation of this structure is that $\u$ describes the microscopic dynamics of individual oscillators, while their effective damping, $a_1$, stiffness, $a_0$, and forcing, $g$, may depend on a collective  quantity, such as a mean field of the form $U(\u)= N^{-1}\sum_{k=1}^N u_k$.

Assuming a solution $\u$  is known or may eventually be constructed, we may treat the coefficients $a_0$ and $a_1$ as explicit functions of time.
In this light, we expect that each second order ODE in \eqref{eq:microscopic_sec_order} admits two linearly independent fundamental solutions of the homogeneous equations, along with a particular solution of the inhomogeneous equations determined by $g$. 
This observation motivates the following procedure.

Our  goal is to find a lower dimensional, closed-form description for the microscopic dynamics, i.e., the evolution of $\u(t)$.
To this end, we posit that a superposition principle exists, and propose the ansatz
\begin{equation}
  \label{eq:ansatz_second_order}
  \u(t) =  \xixi_1 X_1(t) +  \xixi_2 X_2(t) + \onevec Y(t) 
\end{equation}
where $ \onevec =(1,\ldots,1)^T\in\C^N$, and $X_1(t), X_2(t)$ and $Y(t)$ are scalar-valued functions which we refer to as fundamental solutions. Moreover, the ansatz is parameterized by the vectors $\xixi_1\in \C^N$ and $\xixi_2\in\C^N$, the meaning of which will be discussed later.
This superposition mirrors the classical theory of second-order linear ODEs, where the general solution is formed from two homogeneous fundamental solutions plus a particular solution of the inhomogeneous system.

Applying this ansatz to \eqref{eq:microscopic_sec_order},
we obtain $N$ identical equations which --- upon invoking the linear independence of the fundamental solutions $X_1,X_2$ and $Y$--- collapse to $3$ \emph{macroscopic equations},
\begin{subequations}
  \label{eq:macroscopic_sec_order}
  \begin{align}\label{eq:macroscopic_sec_order_a}
    \ddot{X}_1 + \fn{a_1}(\u,t) \dot{X}_1 + \fn{a_0}(\u,t) X_1 &= 0,
    \\\label{eq:macroscopic_sec_order_b}
    \ddot{X}_2 + \fn{a_1}(\u,t) \dot{X}_2 +\fn{a_0}(\u,t) X_2 &= 0,
    \\\label{eq:macroscopic_sec_order_c}
    \ddot{Y} + \fn{a_1}(\u,t) \dot{Y} + \fn{a_0}(\u,t) Y &= \fn{g}(\u,t),
  \end{align}
\end{subequations}
where $\u$ is given by the superposition ansatz in \eqref{eq:ansatz_second_order}, which closes the equations.
% dimensional reduction:
The macroscopic dynamics are governed by three scalar second-order ODEs for the functions $X_1,X_2,Y$, each contributing two degrees of freedom, resulting in a $ 2 \times 3 = 6 $-dimensional system; this constitutes a significant reduction in complexity compared to the original $2N$-dimensional microscopic system~\eqref{eq:microscopic_sec_order}, and is independent of the number of units $N$ in the microscopic system.

We stress that this reduction to a lower dimensional ensemble of ODEs is non-trivial due to the nonlinear dependence of $a_0$, $a_1$, and $g$ on $\u$, and dependence on $t$. 
The validity of this reduction crucially depends on the availability of the fundamental solutions for the (quasi-linear) macroscopic equations \eqref{eq:macroscopic_sec_order}, while the ansatz \eqref{eq:ansatz_second_order} implicitly assumes that the microscopic system admits a superposition principle.
Fortunately, under mild regularity assumptions on the functions $a_0$, $a_1$ and $g$ --- for instance, Lipschitz continuity in $\u$ --- existence and uniqueness of solutions to the macroscopic equations \eqref{eq:macroscopic_sec_order} follows from the Picard-Lindelöf theorem.
Therefore, even if closed-form expressions for the fundamental solutions as given by \eqref{eq:macroscopic_sec_order} are inaccessible, the macroscopic system can be solved numerically, offering a tractable and efficient representation of the full microscopic dynamics.
Thus, the ansatz yields a self-consistent, dimensionally reduced flow for the system.

A few comments are in place regarding the role of the parameters $\xixi_1$ and $\xixi_2$.
First, note that the $\xixi_1$ and $\xixi_2$ in the superposition ansatz~\eqref{eq:ansatz_second_order} relate solutions of the reduced 6-dimensional macroscopic equations with solutions of the original $2N$-dimensional equations. The ensuing reduction in dimensionality prompts that $2N-6$ quantities --- i.e., constants of motions -- are preserved by the flow. This suggests that the microscopic solution space is densely foliated, structured by invariant manifolds on which dynamics occur. The parameters $\xixi_1$ and $\xixi_2$ choose on which of the manifolds the dynamics occur.
% 
% Canonical choice 
Second, suppose we wish to solve the microscopic system with specified initial data, 
we let the macroscopic initial data be:
\begin{subequations}
  \label{eq:order2-reduced-init-conds}
  \begin{align}
    X_1(0) &= 1\,,\quad \dot{X}_1(0) = 0,
    \\
    X_2(0) &= 0\,,\quad \dot{X}_2(0) = 1,
    \\
    Y(0) &= 0\,,\quad \dot{Y}(0) = 0 .
  \end{align}
\end{subequations}
Recalling the superposition ansatz \eqref{eq:ansatz_second_order}, we then have
\begin{subequations}
  \begin{align}
    \fn{\u}[0] &= \xixi_1 , 
    \fn{\dot{\u}}[0] = \xixi_2.
  \end{align}
\end{subequations}
In this case parameters $\xixi_k$ correspond to initial data of the microscopic system.

Note that the parameters $\xixi_1$ and $\xixi_2$ play a dual role: they serve as parameters for both initial conditions of the microscopic equations~\eqref{eq:microscopic_sec_order} and the coefficients in the ansatz \eqref{eq:ansatz_second_order}, so that the macroscopic dynamics \eqref{eq:macroscopic_sec_order} indirectly depend on these parameters as well.
This dependence is discussed further in Sec.~\ref{sec:special-global-dependence}. 

% Dimensional relationship of parameters:
While the macroscopic system evolves in only six dimensions, the full $2N$-dimensional solution space of the microscopic system is retained through the parameters $\xixi_1,\xixi_2\in \C^N$, which encode the initial conditions of the microscopic dynamics.
These vectors are fixed for each solution and remain invariant under the macroscopic flow, thereby parametrizing the microscopic solution manifold.
In this sense, they act in some sense as generalized constants of motion: while not conserved quantities in the classical dynamical sense, they uniquely label individual solution trajectories and are preserved by the reduced evolution.

% ================================================================

\section{Generalization to higher order quasi-linear systems \label{sec:general-result}}

The dimensional reduction method presented for quasi-linear equations of second order~\eqref{eq:microscopic_sec_order} generalizes naturally to equations involving higher-order derivatives. 
Consider a microscopic system consisting of $N$ ordinary differential equations of order $M$ in the variable $\u: \R \to \C^N$, given by 
\begin{equation}\label{eq:general_microscopic}
  \sum_{j = 0}^{M} \fn{a_j}[\u, t] \der{u}{j}_k = \fn{g}[\u, t]
\end{equation}
for $k = 1, \dots, N$, where $\der{u}{j}_k$ denotes the $j$:th derivative in $t$.
Defining the  differential operator
\begin{equation}
  \fn{L}[\v] = \sum_{j = 0}^{M} \fn{a_j}[\v, t] \diff[j]{}{t}
\end{equation}
allows us to write the microscopic equations in compact form,
\begin{equation} \label{eq:general-original}
  \mathop{\fn{L}[\u]} [ u_k ] = \fn{g}[\u, t],
\end{equation}
for $k = 1, \dots, N$.
This microscopic system has $N$ equations of order $M$, corresponding to a dimensionality of the solutions space of order $N M$.

To construct a macroscopic system, we use the generalized superposition ansatz 
\begin{equation} \label{eq:general_superposition_ansatz}
  \u(t) = \xixi_1 X_1(t) + \dots + \xixi_M X_{M}(t)  + \onevec Y(t) ,
\end{equation}
where $X_\ell : \R \to \C^M$, $\ell=1,\ldots,M$, $Y : \R \to \C$ are scalar functions, $\onevec$ denotes the $N$-dimensional unit vector, and the with parameters $\xixi_\ell\in\C^N$ (see Sec.~\ref{sec:motivating-example} for a brief explanation concerning the role of these parameters). 

In analogy to the reduction for the second-order quasi-linear ensemble, this ansatz reduces the microscopic equations  \eqref{eq:general-original} into a set of $N$ identical equations containing the scalar functions $X_\ell$ and $Y$.
Substituting the superposition ansatz into \eqref{eq:general-original}, invoking  (quasi-)linearity of $\fn{L}$ and linear independence of the fundamental solutions $(X_\ell)$ and $Y$, these equations collapse to the macroscopic system becomes 
\begin{subequations} \label{eq:general_macroscopic}
  \begin{align}
    \mathop{\fn{L}[\u]} [ X_\ell ] &= 0,\quad \ell = 1, \dots, M, \\
    \mathop{\fn{L}[\u]} [ Y ] &= \fn{g}[\u, t].
  \end{align}
\end{subequations}
where $\u$ will be replaced by the superposition ansatz in \eqref{eq:general_superposition_ansatz}.
The initial conditions of the microscopic equations can be prescribed in a clear way by choosing the following initial conditions for the reduced variables:
\begin{subequations} \label{eq:general_macroscopic-ivs}
  \begin{align}
    \der{X}{j}_\ell(0) &= \delta_{j\ell}, \\
    \label{eq:general_macroscopic-ivs-2}
%     \der{X}{j}_j(0) &= 1, \\
    \der{Y}{j}(0) &= 0,
  \end{align}
\end{subequations}
for $j = 0, \dots, M-1$.
The solutions of the microscopic system are then determined by the initial conditions
\begin{equation} \label{eq:general-original-ivs}
  \der{\u}{j}(0) = \xixi_{j+1},\quad j = 0, \dots, M-1,
\end{equation}
where $\xixi_\ell \in \C^N$ act as the initial data.

Given the fundamental functions $(X_j)$ and $Y$ are accessible, they form a set of $M+1$ (linearly independent) fundamental solutions of the macroscopic system.
These fundamental solutions act as parameters of the superposition principle and generate the microscopic flow, $\u(t)$.
Furthermore, the microscopic initial data $\xixi_\ell$ are (constant) parameters preserved along the generated orbits.

A few comments are in place.
The macroscopic system \eqref{eq:general_macroscopic} has dimension $M(M+1)$ --- observe that this dimension is \emph{independent} of the microscopic number of variables, $N$.
Therefore, the macroscopic system has a drastically reduced dimension compared to the microscopic system.
Furthermore, if $\fn{g} \equiv 0$, the inhomogeneous term vanishes and $Y \equiv 0$, the system is closed in $(X_j)$ alone.
We emphasize that this reduction remains valid even when coefficients $a_j$ and the forcing term $g$ include higher order derivatives of $\u$. What is essential for the dimensional reduction to be valid is that all coefficients $a_j$ and $g$ are the same for all units $k$.

% ================================================================

\section{Weighted linear mean field and rescaled macroscopic equations\label{sec:special-global-dependence}}

In  the following, we consider that the coefficients in the macroscopic equations $a_j$~\eqref{eq:general_macroscopic} may depend on the microscopic variables $\u$ via a  function, $U=U(\u)$ with $U:\C^N \to \C$, so that the coefficient functions are defined as  $a_j=a_j(U,t)$.  
While many choices for $U$ are viable,
we focus here on the important case of a scalar-valued function representing a mean-field, such as the weighted linear combination of the microscopic variables $(u_k)$,
\begin{equation}\label{eq:weighted_meanfield}
  U(\u) = \sum_{k=1}^N w_k u_k = \vec{w} \cdot \u.
\end{equation}

Although the dimensional reduction discussed further above reduces the dimensionality of the microscopic system with degrees of freedom of order $NM$ to $M(M + 1)$ of the macroscopic system, the macroscopic equations are still expressed in terms of the $  N  M $ parameters controlling the initial conditions in \eqref{eq:general-original-ivs}.
However, the choice of a linearly weighted mean field dependence on $ \u $ reduces the number of parameters further.

To illustrate this reduction in a concrete case, let us specialize to the second-order system introduced earlier in~\eqref{eq:microscopic_sec_order}, i.e., with $M = 2$ and $a_2 \equiv 1$.
If coefficients $a_i$ only depend on the weighted sum of individual states in~\eqref{eq:weighted_meanfield},
the macroscopic system has the form:
\begin{subequations}\label{eq:U_equations}
\begin{align}
  \ddot{X}_1 + a_1(U) \dot{X}_1 + a_0(U) X_1 = 0,\\
  \ddot{X}_2 + a_1(U) \dot{X}_2 + a_0(U) X_2 = 0,\\
  \ddot{Y} + a_1(U) \dot{Y} + a_0(U) Y = g(U).
\end{align}
\end{subequations}
Expressing the weighted mean field $U$ in terms of the fundamental solutions $X_1,X_2,Y$, using the superposition~\eqref{eq:ansatz_second_order}, 
\begin{align}\label{weighted_U}
  \begin{split}
    U &=  (\vec{w} \cdot \xixi_1) X_1 + (\vec{w} \cdot \xixi_2) X_2  +  (\vec{w} \cdot \onevec) Y
  \end{split}
\end{align}
closes the system of equations~\eqref{eq:U_equations}.
Note that, while the mean field depends on the weights $\w$ and initial data $\xixi_1$ and $\xixi_2$, it effectively depends only on the three scalar parameters given by $ \vec{w} \cdot \xixi_1 $, $ \vec{w} \cdot \xixi_2 $ and $ \vec{w} \cdot \onevec $, which capture all influence of the microscopic initial conditions on the macroscopic dynamics through the weighted mean field. 

To remove explicit dependence on the microscopic initial data from the macroscopic system, we define rescaled variables that absorb these parameters into the dynamical variables themselves.
Specifically, quasi-linear structure of the ODEs and the linear superposition principle allow to cast the macroscopic equations into parameter-free form, by introducing the rescaled variables
\begin{align}
    \begin{split}
    \tilde X_1 &= (\vec{w} \cdot \xixi_1) X_1,\\
    \tilde X_2 &= (\vec{w} \cdot \xixi_2) X_2, \\
    \tilde Y &= (\vec{w} \cdot \onevec) Y.
    \end{split}
\end{align}
In terms of the rescaled variables, the weighted mean field becomes
\begin{equation}
U = \tilde X_1 + \tilde X_2 + \tilde Y,
\end{equation}
and the rescaled macroscopic equations read
\begin{subequations}\label{eq:U_equations_rescaled}
\begin{align}
  \ddot{\tilde X}_1 + a_1(U) \dot{\tilde X}_1 + a_0(U) \tilde X_1 &= 0,\\
  \ddot{\tilde X}_2 + a_1(U) \dot{\tilde X}_2 + a_0(U) \tilde X_2 &= 0,\\
  \ddot{\tilde Y} + a_1(U) \dot{\tilde Y} + a_0(U) \tilde Y &= \tilde g(U),
\end{align}
\end{subequations}
where we  introduced the rescaled forcing $\tilde  g = (\vec{w} \cdot \onevec) g $.

Two points deserve being noted. First, the extension of this approach to quasi-linear systems of higher order ($M>2$) (see Sec.~\ref{sec:general-result}) is straightforward.
Second, note that the mean-field does not need to sum over all $u_k$, one may choose $w_k=0$ for some $k$.

% ================================================================

\section{Example systems\label{sec:examples}}
\subsection{Circuit with nonlinear capacity\label{sec:example1}}
Here we apply our formalism to an example on electronic circuits. 
We consider an ensemble of $N$ series RLC circuits. The current $u$ for an isolated RLC circuit follows this ODE:
\begin{equation}
\ddot{u} + \frac{R}{L}\dot{u}+\frac{1}{LC}u = 0
\end{equation}
where $R,L,C$ are the resistance, inductance and capacitance of the circuit. 

We couple the ensemble through voltage-controlled capacitors.
For example, consider replacing the capacitors with varactor diodes that are controlled by the average current of the entire ensemble.
We model the capacitance of the varactor diode with respect to applied voltage $V$,
 \begin{equation}
C(V) = \frac{C_0}{(1+V/V_0)^{1/2}}
\end{equation}
where $C_0$ and $V_0$ are parameters of the specific diode design. 
Now suppose the voltage controlling the capacitances is induced by the average current: $V = R_0 U$, 
where $U = 1/N\sum\limits_{k=1}^N u_k$ is the average current. 

An ensemble of $N$ coupled RLC circuits then evolves according to:
\begin{equation}\label{eq:ex1_microscopic}
\ddot{u}_k + a \dot{u}_k + \sqrt{b+cU} u_k = 0
\end{equation}
where ${\bm u}=(u_1,\ldots,u_N)$ are the currents of the RLC circuit, and constants $a,b,c$ are determined by their components:
$a = RL^{-1}$, $b = (LC_0)^{-2}$, $c = R_0V_0^{-1}(LC_0)^{-2}$. 
This systems is of the form~\eqref{eq:microscopic_sec_order}. 
All quantities are real and coefficients $a,b,c$ are all non-negative.

\paragraph{Dimensional reduction.}
Since all quantities are real and the right-hand-side of our example~\eqref{eq:ex1_microscopic} is zero, this system will follow 4-dimensional dynamics. 

We write the microscopic variables $u_k$ in terms of the fundamental solutions $X_1$, $X_2$ and the initial conditions,
\begin{equation}\label{eq:ex1_superposition}
u_k(t) =  (\xixi_1)_k X_1(t) + (\xixi_2)_k X_2(t).
\end{equation}
These macroscopic variables evolve according to
\begin{subequations}\label{eq:ex1_macroscopic}
\begin{align}
\ddot{X}_1 + a \dot{X}_1 + \sqrt{b+c\,U}X_1 &= 0,\\
\ddot{X}_2 + a \dot{X}_2 + \sqrt{b+c\,U}X_2 &= 0,
\end{align}
\label{example_X}
\end{subequations}
with initial conditions as defined by~\eqref{eq:order2-reduced-init-conds}.
To close \eqref{eq:ex1_macroscopic}, we express the average current $U$ in terms of $X_1$ and $X_2$:
\begin{equation}
\begin{split}
U 
&= \frac{1}{N}\sum\limits_{k=1}^N u_k \\
&= \frac{1}{N}\sum\limits_{k=1}^N \Big((\xixi_1)_k X_1+(\xixi_2)_k X_2 \Big) \\
&= \bar{\xi}_1 X_1(t) + \bar{\xi}_2 X_2(t) \\
\end{split}
\end{equation}
where we define $\bar{\xi}_\ell = 1/N \sum_k (\xixi_\ell)_k$.
Substituting this into~\eqref{example_X} we get two nonlinear second-order ODEs,
\begin{subequations}
\begin{align}\label{eq:ex1_macroscopic_explicit}
\ddot{X}_1 + a \dot{X}_1 + \sqrt{b+c\left( \bar{\xi}_1 X_1 +  \bar{\xi}_2 X_2 \right)}X_1 &= 0,\\
\ddot{X}_2 + a \dot{X}_2 + \sqrt{b+c\left(\bar{\xi}_1 X_1 + \bar{\xi}_2 X_2 \right)}X_2 &= 0,
\end{align}
\label{example_closed_eqs}
\end{subequations}
\noindent which solve the microscopic system~\eqref{eq:ex1_microscopic} via the  superposition~\eqref{eq:ex1_superposition} in closed form.

\begin{figure}
  \begin{overpic}[width=0.44\textwidth]{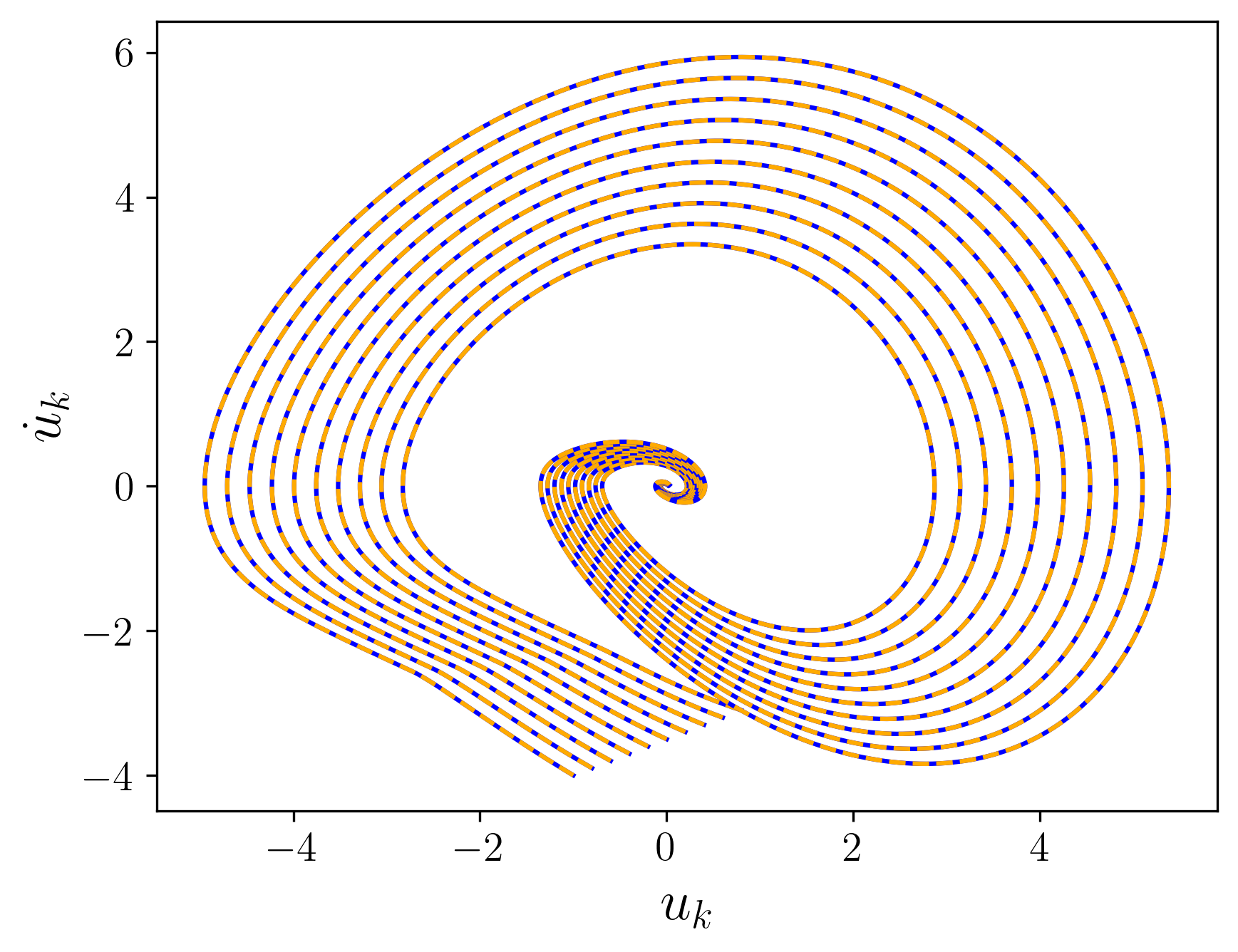}
  \end{overpic}
  \caption{
    Comparison of microscopic dynamics obtained by direct integration of the microscopic equations~\eqref{eq:ex1_microscopic} (blue); and by integrating macroscopic equations~\eqref{eq:ex1_macroscopic_rescaled} whose solutions generate the microscopic solutions via the superposition ~\eqref{eq:ex1_superposition} (orange dashed).
    Parameters are $N=10, a=0.6,b=0.5,c=0.3$.
  }
  \label{fig:example1}
\end{figure}

\paragraph{Rescaled macroscopic equations.}
To remove explicit dependence on the microscopic initial data from the macroscopic system, we introduce the rescaled variables,
\begin{equation}
\tilde X_1 = \bar{\xi}_1 X_1\,, \quad \tilde X_2 = \bar{\xi}_2 X_2.
\end{equation}
which are satisfied by the rescaled macroscopic equations, 
\begin{subequations}
\begin{align}\label{eq:ex1_macroscopic_rescaled}
\ddot{\tilde X}_1 + a \dot{\tilde X}_1 + \sqrt{b+c\left( \tilde X_1 + \tilde X_2 \right)}\tilde X_1 &= 0,\\
\ddot{\tilde X}_2 + a \dot{\tilde X}_2 + \sqrt{b+c\left( \tilde X_1 + \tilde X_2 \right)}\tilde X_2 &= 0,
\end{align}
\end{subequations}
with initial conditions,
\begin{subequations}
\begin{align}
\tilde X_1(0) &= \bar{\xi}_1\,, \quad &\dot{\tilde X}_1(0) &= 0,\\
\tilde X_2(0) &= 0\,, \quad &\dot{\tilde X}_2(0) &= \bar{\xi}_2.
\end{align}
\end{subequations}
Numerically solving the rescaled macroscopic equations~\eqref{eq:ex1_macroscopic_rescaled}, we may recover the exact microscopic dynamics of~\eqref{eq:ex1_microscopic}, as illustrated in Fig.~\ref{fig:example1}. 
The microscopic variables are retrieved via
\begin{align}
 u_k(t) &= \frac{(\xixi_1)_k}{\bar{\xi}_1}\tilde{X}_1(t) + \frac{(\xixi_2)_k}{\bar{\xi}_2}\tilde{X}_2(t).
\end{align}
The comparison in Fig.~\ref{fig:example1} illustrates a perfect match.

\subsection{Nonlinearly damped oscillator}
In the following, we consider a more general, complex-valued example. 
Drawing an analogy with the previous electronic example, this could be for instance active circuit with complex impedances. 
In this analogy the ensemble variables $u_k \in \C$ could be analytic current signals, with real part of $u_k$ representing  current and the imaginary part its Hilbert transform. 

An example of such dynamics is 
\begin{equation}
\ddot{u}_k -i \left(1+a\,\text{Im}\left[ \frac{\dot{u}_1}{u_1} \right] \right)\dot{u}_k +b u_k = 0,
\label{eq:example2}
\end{equation}
where each unit is driven by the first unit, $u_1$, rather than some global mean-field --- an example of a simple weighted linear mean field (Sec.~\ref{sec:special-global-dependence}).
Initial conditions are defined by \eqref{eq:order2-reduced-init-conds}, so that ${\bm \u}(0) = \xixi_1$ and $\dot{{\bm \u}}(0) = \xixi_2$.

We can check that the energies for each unit,
\begin{equation}
E_k = |\dot{u}_k|^2 + b |u_k|^2
\label{conserved}
\end{equation}
are conserved in time,
\begin{equation}
\begin{split}
\dot{E}_k &= \ddot{u}_k \dot{u}^*_k + b u_k \dot{u}^*_k + \text{c.c.} = \\
&= \dot{u}^*_k (\ddot{u}_k+b u_k) + \text{c.c.} = \\
&= -\lambda |\dot{u}_k|^2 + \text{c.c.} = 0,
\end{split}
\end{equation}
where $\lambda = -i(1+a\,\text{Im}[\dot{u}_1/u_1])$ is the damping term, 
${}^*$ denotes complex conjugation and ``c.c.'' stands for the complex conjugate of all the terms. 
Here, we substituted the governing equations \eqref{eq:example2} and used the property that the damping term is purely imaginary, i.e. $\lambda^* = -\lambda$.

\paragraph{Dimensional reduction.}
The right-hand-side of system~\eqref{eq:example2} is zero, but the quantities $u_k\in\C$ are now complex and so to are the macroscopic variables $X_1,X_2\in\C$ which evolve according to:
\begin{subequations}
\begin{align}
\ddot{X}_1 -i \left(1+a\,\text{Im}\left[ \frac{\dot{u}_1}{u_1} \right] \right) \dot{X}_1 + b X_1 &= 0,\\
\ddot{X}_2 -i \left(1+a\,\text{Im}\left[ \frac{\dot{u}_1}{u_1} \right] \right) \dot{X}_2 + b X_2 &= 0,\
\end{align}
\label{eq:tmp}
\end{subequations}
with the same initial conditions as in~\eqref{eq:order2-reduced-init-conds}. 
The two complex-valued, second order ODEs possess a constant of motion analogous to ~\eqref{conserved}.  
The microscopic variables $u_k$ are given via the superposition of the fundamental solutions $X_1$ and $X_2$,
\begin{equation}
u_k(t) =  (\xixi_1)_k X_1(t)+(\xixi_2)_kX_2(t).
\label{example2_trans}
\end{equation}

As in the previous example, we close the system of equation, which in this case requires expressing the coupling term ${\dot{u}_1}/{u_1}$ in terms of $X_1,X_2$,
\begin{equation}
\frac{\dot{u}_1}{u_1} = \frac{ (\xixi_1)_1\dot{X}_1 +  (\xixi_2)_1\dot{X}_2}{ (\xixi_1)_1X_1 +  (\xixi_2)_1X_2},
\end{equation}
and plugging it into equations~\eqref{eq:tmp} to obtain a closed set:
\begin{subequations}
\begin{align}
\ddot{X}_1 -i \left(1+a\,\text{Im}\left[ \frac{ (\xixi_1)_1\dot{X}_1 +  (\xixi_2)_1\dot{X}_2}{ (\xixi_1)_1X_1 +  (\xixi_2)_1X_2} \right] \right) \dot{X}_1 + b X_1 &= 0,\\
\ddot{X}_2 -i \left(1+a\,\text{Im}\left[ \frac{ (\xixi_1)_1\dot{X}_1 +  (\xixi_2)_1\dot{X}_2}{ (\xixi_1)_1X_1 +  (\xixi_2)_1X_2} \right] \right) \dot{X}_2 + b X_2 &= 0.
\end{align}
\end{subequations}
This system, together with relation~\eqref{example2_trans} exactly solves system~\eqref{eq:example2}. 

\begin{figure}[h]
  \includegraphics[width=0.44\textwidth]{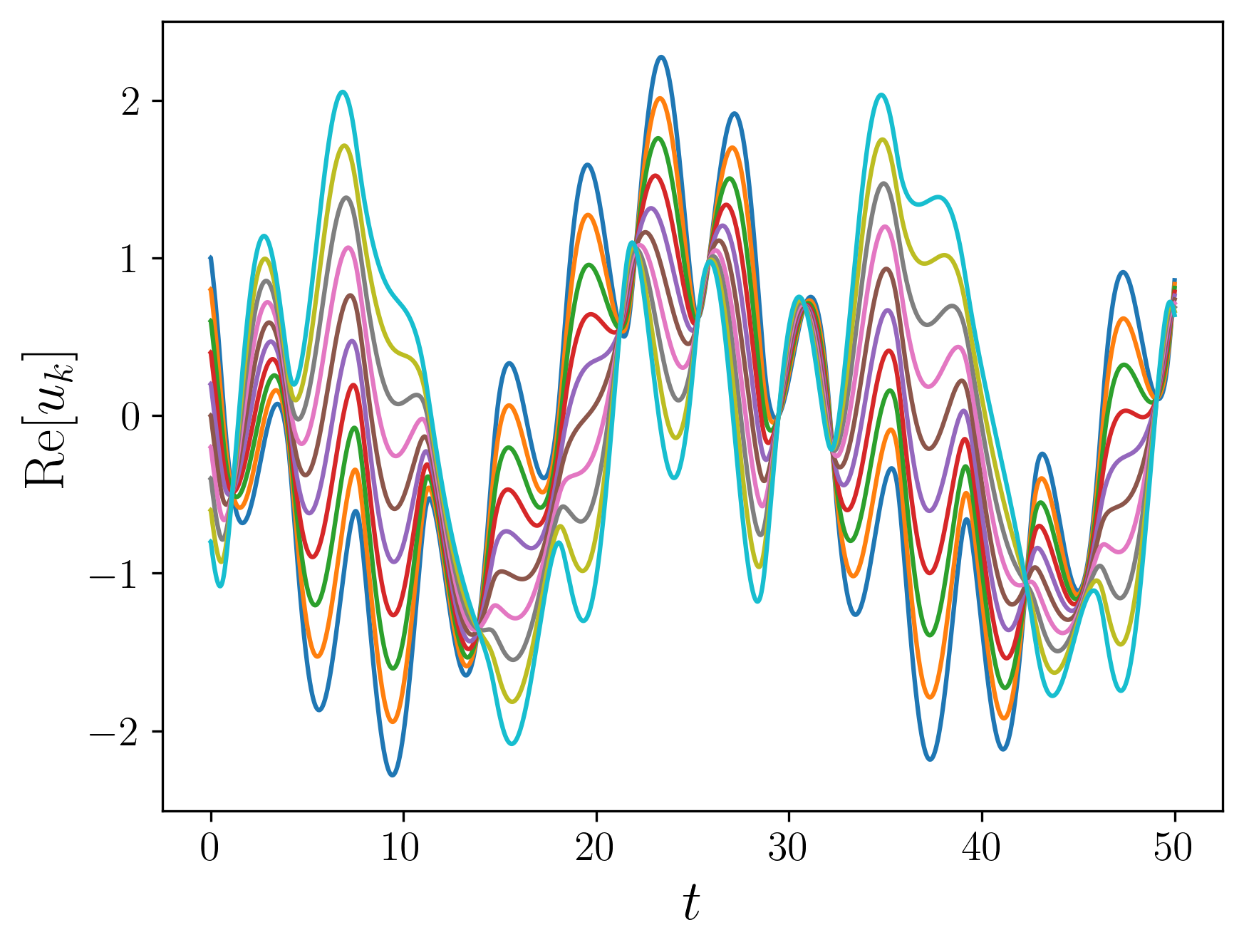}\\
  \includegraphics[width=0.44\textwidth]{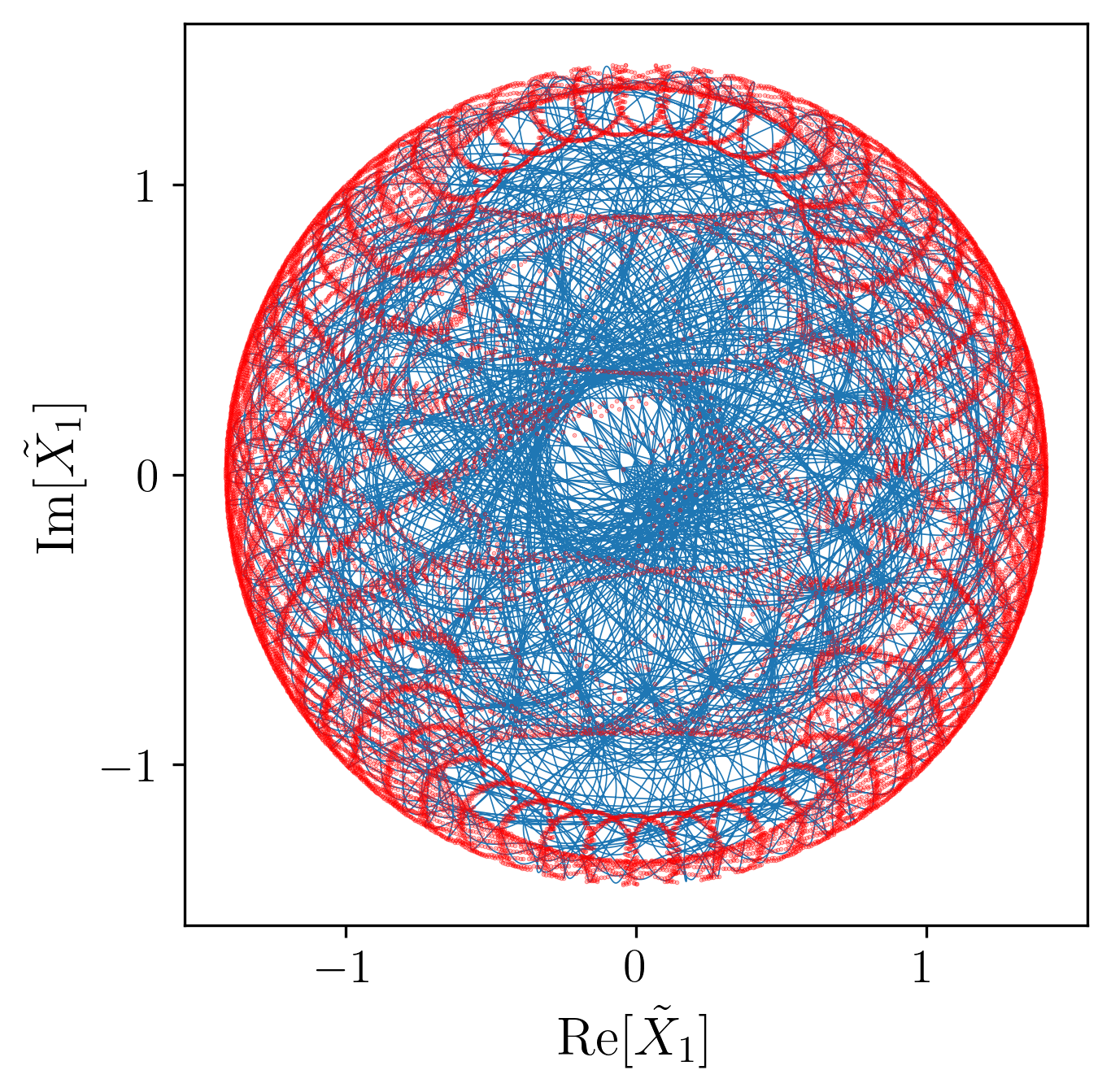}
  \caption{
    System~\eqref{eq:example2} for $a=-0.7$ and $b=0.5$ and $N=10$.
    In the top panel the real parts of $u_j$ in time (representing currents in the electronic analogy), and in the bottom panel the $X_1(t)$ (blue curves) in the complex plane alongside with the Poincaré section when $X_2$ crosses $\text{Im}(X_2)=0$ (red dots).
    For the chosen parameters and initial conditions the system exhibits chaotic dynamics as indicated by a positive maximal Lyapunov exponent. %$\lambda_\text{max}\approx 0.0507$
    $(\xixi_1)_1 = 1-i,(\xixi_2)_1 = i-1$ with initial conditions given by \eqref{eq:order2-reduced-init-conds}. 
  }
  \label{fig:example2}
\end{figure}
\paragraph{Rescaled macroscopic equations.}
Defining the rescaled variables,
\begin{equation}
\tilde X_1 = (\xixi_1)_k X_1\,, \quad \tilde X_2 = (\xixi_2)_kX_2,
\end{equation}
the parameter free macroscopic equations become:
\begin{subequations}
\begin{align}
\ddot{\tilde X}_1 -i \left(1+a\,\text{Im}\left[ \frac{\dot{\tilde X}_1 + \dot{\tilde X}_2 }{\tilde X_1 + \tilde X_2 } \right] \right) \dot{\tilde X}_1 + b\, \tilde X_1 &= 0.\\
\ddot{\tilde X}_2 -i \left(1+a\,\text{Im}\left[ \frac{\dot{\tilde X}_1 + \dot{\tilde X}_2}{\tilde X_1 + \tilde X_2 } \right] \right) \dot{\tilde X}_2+ b\, \tilde X_2 &= 0.
\end{align}
\end{subequations}

This example illustrates the dimensional reduction approach for complex-valued coupled systems with nonlinear phase-dependent coefficients, and shows chaotic dynamics captured by the reduced macroscopic equations, as illustrated in Fig.~\ref{fig:example2}.

% ================================================================

\section{Discussion\label{sec:discussion}}

% Summary: 
We describe an exact dimensional reduction for ensembles of quasi-linear ordinary differential equations (ODEs) characterized by higher-order linear differential equations~\eqref{eq:general_microscopic}.
Unlike well-studied models tied to Riccati equations --- such as the Kuramoto, Theta, or Quadratic Integrate-and-Fire models~\cite{Bick2018c} --- our approach applies to a different class of systems,  revealing that their $NM$-dimensional dynamics can be exactly captured by a small set of $M$ macroscopic variables governed by $M+1$ equations, providing the fundamental solutions that parameterize the superposition reconstructing the full microscopic dynamics.
This reduction not only provides access to simplified analysis via the macroscopic equations~\eqref{eq:general_macroscopic}, but also offers the practical means to reduce computational complexity (see examples in Sec.~\ref{sec:examples}).

% Implications:
The dimensional reduction offers two advantages:
First, it provides simpler access analysis for otherwise intractable systems, offering insight into stability, bifurcations, and collective dynamics;
second, the reduction facilitates efficient numerical methods, allowing the reconstruction of microscopic states from very low-dimensional dynamics, independent of the network size $N$ --- this is similar to the Watanabe-Strogatz method\cite{watanabe1994} or the analogous generation of orbits via Möbius group actions\cite{marvel2009,chen2017} for phase oscillator ensembles. 
Extending this perspective to higher-order quasi-linear equations suggests a rich avenue for discovering new solvable nonlinear systems, analogous to the relation between second-order linear dynamics and systems of Riccati-like equations\cite{inceordinary,CestnikMartens2024} (see App.~\ref{sec:riccati-linkage}). 

% Potential for diverse applications:
Since the coupling functions $a_j(\u,t)$ in \eqref{eq:general_microscopic} can be expressed in terms of the reduced variables, ensembles of quasi-linear ODEs \eqref{eq:microscopic_sec_order}  apply naturally to physical systems, such as coupled mechanical oscillators or nonlinear electronic LC circuits, where each unit follows the same linear dynamics but interacts through nonlinear mean field interactions, often related to collective modes or energy transfer pathways. 
Thus, our results illuminate how physically relevant systems can exhibit emergent low-dimensional behavior, setting the stage for further exploration of nonlinear network dynamics. 

A related mathematical structure appears in the time-independent Schrödinger equation, as it involves second-order linear ODEs; however, the analogy is limited since quantum mechanical problems typically impose boundary conditions to solve for eigenmodes, rather than initial conditions, making the notion of evolving microscopic degrees of freedom with arbitrary initial data inapplicable.
More concretely, boundary conditions are usually globally imposed by the potential so in this context it is unnatural to think of different boundary conditions.
Still, this comparison displays the generality of the framework in describing wave-like dynamics with structured interactions.

% limitations: 
The dimensional reduction presented here applies specifically to ensembles of quasi-linear ODEs with identical structure, i.e., the coupling to mean field forcings occurs through the identical coefficient functions $a_j$. 
In general we cannot expect fundamental solutions of the system to be analytically available. 
Nevertheless, even in such cases, the dimensional reduction remains valuable, since numerically computing the reduced macroscopic system is much more efficient, while at the same it enables analytical tools, such as bifurcation analysis that would be intractable at the microscopic scale.

% GENERALIZATIONS
Just like there is a nonlinear transformation between second order quasi-linear ensembles and Riccati-like ensembles, as explained in the Appendix~\ref{sec:riccati-linkage}, it may be possible to find other transformations that map the quasi-linear structure into some other nonlinear systems of practical interest.
Future research should explore such possibilities.
Such mappings could reveal broader classes of nonlinear dynamics that admit similar reductions.
A further possibility is to invert the reduction: one might design microscopic dynamics that produce a prescribed evolution for the mean-field $U$. 
Note that, in general, there are multiple ways to design micro-dynamics producing the same mean field.
This can for example be seen by considering the linear weighted mean field with arbitrary weight coefficients.
Finally, the (quasi-)linear structure may extend naturally from scalars to vector- or matrix-valued systems.
Similar generalizations have been carried out for Riccati-type equations involving vector and matrix variables.\cite{Lohe18,Lohe19}

% Future work:
For future work, it would be interesting to explore possibilities for a description of the dynamics in the continuum limit with infinitely many oscillators, $N\to\infty$, similar to classical work for the Kuramoto model and related Riccati-like dynamics.\cite{mirollo1991stability,Bick2018c,PazoCestnik25}
It would be interesting to investigate if invariant manifolds like the Ott-Antonsen manifold\cite{ott2008} exist for special choices of the quasi-linear system.
Notably, we know that in the case the case of second order quasi-linear ODEs a mapping to Riccati-like systems is feasible (App.~\ref{sec:riccati-linkage}), at least for finite size systems. However, note that the correspondence between finite size and continuum limit descriptions in terms of probability density functions is not one to one which may complicate matters.

Moreover, the dimensional reduction does not immediately generalize to systems with strong heterogeneity, stochastic forcing, or delay.
Further research should explore the possibility to find reductions for heterogeneous units in the continuum limit, much like like.\cite{OA08,LA15,Laing15,CestnikPikovsky2022,CestnikPikovsky2022b,PietrasPikovsky2023,PazoCestnik25}
Finally, many systems evolve in the presence of noise.
Indeed, stochastic forcing terms can be applied via the coefficient functions $a_j$ and the forcing $g$, provided that dynamics is subject to identical forcing for all oscillators $k$.
This contrasts stochastic forcing terms, $\eta_k(t)$, acting on an individual level, e.g., $\ddot{u}_k + a_1 \dot{u}_k +a_0 u_k = g + \eta_k(t)$.
It may be possible to treat such noise in a similar way as in previous studies.\cite{ToenjesPikovksy2020,PietrasPikovsky2023,ClusellaMontbrio2024}

\begin{acknowledgments}
    We gratefully acknowledge financial support from the Royal Swedish Physiographic Society of Lund.
%     We are grateful to Sanjay Dharmavaram for helpful comments and insightful discussions.
\end{acknowledgments}

\appendix

\section{Relationship between Riccati-like equations and second order quasi-linear equations}\label{sec:riccati-linkage}

We explain here the relationship between Riccati-like and second order quasi-linear equations, which builds upon a well known transformation between Riccati equations and second order linear ordinary differential equations.\cite{inceordinary}

Consider the nonlinear Riccati system
\begin{equation} \label{eq:basic-nonlinear-riccati}
  \dot{x}_k = \fn{a}[t] x_k^2 + \fn{b}[\vec{x}, t] x_k + \fn{c}[\vec{x}, t]
  .
\end{equation}
This is a less general version of the system in Ref.~\cite{CestnikMartens2024}, where $ \fn{a} $ can be a function of $ \vec{x} $;
here, $ a $ is only a function of time $ t $ to simplify the following calculations.
We will show that through a simple transformation, there is a natural correspondence between the reduction \cref{eq:general_superposition_ansatz} and the reduction
\begin{equation} \label{eq:cestnik-martens-reduction}
  x_k = Q + \frac{y \xi_k}{1 + s \xi_k}
\end{equation}
used in Ref.~\cite{CestnikMartens2024}. Here constants $\xi_k = x_k(0)$ are not to be confused with our constants $\xixi_j$ defined in Eq.~\eqref{eq:general-original-ivs}, even though they are related. The reduced variables have the dynamics
\begin{subequations} \label{eq:cestnik-martens-reduction-dynamics}
  \begin{align}
    \dot{Q} &= \fn{a}[t] Q^2 + \fn{b}[Q, y, s, t] Q + \fn{c}[Q, y, s, t]
    \\
    \dot{y} &= \left( \fn{b}[Q, y, s, t] + 2 \fn{a}[t] Q \right) y
    \\
    \dot{s} &= - \fn{a}[t] y
    ,
  \end{align}
\end{subequations}
with $ \fn{b} $ and $ \fn{c} $ having natural definitions given by \cref{eq:cestnik-martens-reduction}.

If \cref{eq:basic-nonlinear-riccati} is transformed into a system in $ \u $ by the transformation
\begin{equation}
  \label{eq:riccati-transformation}
  x_k = - \frac{\dot{u}_k}{\fn{a}[t] u_k}
  ,
\end{equation}
the new system will have second order dynamics
\begin{equation} \label{eq:transformed-riccati-system}
  \ddot{u}_k + \fn{a_1}[\u, \dot{\u}, t] \dot{u}_k + \fn{a_0}[\u, \dot{\u}, t] u_k = 0
\end{equation}
where the functions $ \fn{a_1}, \fn{a_0} $ correspond to $ \fn{a}, \fn{b}, \fn{c} $ through the transformation
\begin{align}
  \fn{a_1}[\u, \dot{\u}, t] &= - \fn{b}[- \dot{\u} \oslash (\fn{a}[t] \u), t] - \frac{\fn{\dot{a}}[t]}{\fn{a}[t]}
  \\
  \fn{a_0}[\u, \dot{\u}, t] &= \fn{a}[t] \cdot \fn{c}[- \dot{\u} \oslash (\fn{a}[t] \u), t]
  ,
\end{align}
where $ \oslash $ is element-wise division.
The initial conditions
\begin{equation}
  x_k(0) = \xi_k
\end{equation}
of the original problem will translate into the second order one via
\begin{equation}
  \frac{\fn{\dot{u}_k}[0]}{\fn{a}[0] \fn{u_k}[0]} = - \xi_k
  .
\end{equation}
This gives us a choice in how to pick initial conditions for \( \u \) and \( \dot{\u} \), but we can choose $ u_k(0) = 1 $, $ \dot{u}_k(0) = -\fn{a}[0] \xi_k $ without any loss of generality.
Using our main result we then have a reduced system in $ X_1, X_2 $ which relates to our original system like
\begin{equation}
  x_k = -\frac{\dot{X}_1 - \xi_k \fn{a}[0] \dot{X}_2}{\fn{a}[t] \left( X_1 - \xi_k \fn{a}[0] X_2 \right)}
  .
\end{equation}
Note that the reduction variable $ Y $ is not needed, since $ g \equiv 0 $ in \cref{eq:transformed-riccati-system}.
We can relate this reduction to \cref{eq:cestnik-martens-reduction} to get that
\begin{align}\label{eq:Qys}
  \begin{split}
    \frac{Q + (Q s + y) \xi_k}{1 + s \xi_k} &=
    Q + \frac{y \xi_k}{1 + s \xi_k}
    \\
    &= -\frac{\dot{X}_1 - \xi_k \fn{a}[0] \dot{X}_2}{\fn{a}[t] \left( X_1 - \xi_k \fn{a}[0] X_2 \right)}
    \\
    & = \frac{-\frac{\dot{X}_1}{\fn{a}[t] X_1} + \frac{\fn{a}[0]}{\fn{a}[t]} \frac{\dot{X}_2}{X_1} \xi_k}{1 - \fn{a}[0] \frac{X_2}{X_1} \xi_k}
    ,
  \end{split}
\end{align}
and therefore
\begin{subequations}
  \begin{align}
    Q &= - \frac{\dot{X}_1}{\fn{a}[t] X_1}
    \\
    y &= \frac{\fn{a}[0]}{\fn{a}[t]} \left( \frac{\dot{X}_2}{X_1} - \frac{\dot{X}_1}{X_1} \frac{X_2}{X_1} \right)
    \\
    s &= - \fn{a}[0] \frac{X_2}{X_1}
    .
  \end{align}
\end{subequations}
Note that the dynamics \eqref{eq:cestnik-martens-reduction-dynamics} of $ Q $, $ y $ and $ s $ could therefore have been calculated purely from the dynamics in $ X_1 $ and $ X_2 $. Note, that we were able to express $Q,y,s$ with only two variables $X_1,X_2$ due to assuming $a(t)$ only depends on time, otherwise, if $a$ also depends on the system variables then equations~\eqref{eq:Qys} are not reducible.\cite{CestnikMartens2024}

\section*{References}
% \bibliography{references.bib}

\begin{thebibliography}{27}%
\makeatletter
\providecommand \@ifxundefined [1]{%
 \@ifx{#1\undefined}
}%
\providecommand \@ifnum [1]{%
 \ifnum #1\expandafter \@firstoftwo
 \else \expandafter \@secondoftwo
 \fi
}%
\providecommand \@ifx [1]{%
 \ifx #1\expandafter \@firstoftwo
 \else \expandafter \@secondoftwo
 \fi
}%
\providecommand \natexlab [1]{#1}%
\providecommand \enquote  [1]{``#1''}%
\providecommand \bibnamefont  [1]{#1}%
\providecommand \bibfnamefont [1]{#1}%
\providecommand \citenamefont [1]{#1}%
\providecommand \href@noop [0]{\@secondoftwo}%
\providecommand \href [0]{\begingroup \@sanitize@url \@href}%
\providecommand \@href[1]{\@@startlink{#1}\@@href}%
\providecommand \@@href[1]{\endgroup#1\@@endlink}%
\providecommand \@sanitize@url [0]{\catcode `\\12\catcode `\$12\catcode
  `\&12\catcode `\#12\catcode `\^12\catcode `\_12\catcode `\%12\relax}%
\providecommand \@@startlink[1]{}%
\providecommand \@@endlink[0]{}%
\providecommand \url  [0]{\begingroup\@sanitize@url \@url }%
\providecommand \@url [1]{\endgroup\@href {#1}{\urlprefix }}%
\providecommand \urlprefix  [0]{URL }%
\providecommand \Eprint [0]{\href }%
\providecommand \doibase [0]{http://dx.doi.org/}%
\providecommand \selectlanguage [0]{\@gobble}%
\providecommand \bibinfo  [0]{\@secondoftwo}%
\providecommand \bibfield  [0]{\@secondoftwo}%
\providecommand \translation [1]{[#1]}%
\providecommand \BibitemOpen [0]{}%
\providecommand \bibitemStop [0]{}%
\providecommand \bibitemNoStop [0]{.\EOS\space}%
\providecommand \EOS [0]{\spacefactor3000\relax}%
\providecommand \BibitemShut  [1]{\csname bibitem#1\endcsname}%
\let\auto@bib@innerbib\@empty
%</preamble>
\bibitem [{\citenamefont {Kuramoto}(1984)}]{kuramoto1984}%
  \BibitemOpen
  \bibfield  {author} {\bibinfo {author} {\bibfnamefont {Y.}~\bibnamefont
  {Kuramoto}},\ }\href@noop {} {\emph {\bibinfo {title} {{Chemical
  Oscillations, Waves, and Turbulence}}}}\ (\bibinfo  {publisher} {Springer},\
  \bibinfo {address} {Berlin},\ \bibinfo {year} {1984})\BibitemShut {NoStop}%
\bibitem [{\citenamefont {Strogatz}(2000)}]{strogatz2000}%
  \BibitemOpen
  \bibfield  {author} {\bibinfo {author} {\bibfnamefont {S.~H.}\ \bibnamefont
  {Strogatz}},\ }\bibfield  {title} {\enquote {\bibinfo {title} {{From Kuramoto
  to Crawford: exploring the onset of synchronization in populations of coupled
  oscillators}},}\ }\href {\doibase 10.1016/S0167-2789(00)00094-4} {\bibfield
  {journal} {\bibinfo  {journal} {Physica D}\ }\textbf {\bibinfo {volume}
  {143}},\ \bibinfo {pages} {1--20} (\bibinfo {year} {2000})}\BibitemShut
  {NoStop}%
\bibitem [{\citenamefont {Pikovsky}, \citenamefont {Rosenblum},\ and\
  \citenamefont {Kurths}(2003)}]{pikovsky2003}%
  \BibitemOpen
  \bibfield  {author} {\bibinfo {author} {\bibfnamefont {A.}~\bibnamefont
  {Pikovsky}}, \bibinfo {author} {\bibfnamefont {M.}~\bibnamefont {Rosenblum}},
  \ and\ \bibinfo {author} {\bibfnamefont {J.}~\bibnamefont {Kurths}},\
  }\href@noop {} {\emph {\bibinfo {title} {{Synchronization: A Universal
  Concept in Nonlinear Sciences}}}}\ (\bibinfo  {publisher} {Cambridge
  University Press},\ \bibinfo {year} {2003})\BibitemShut {NoStop}%
\bibitem [{\citenamefont {Arenas}\ \emph {et~al.}(2008)\citenamefont {Arenas},
  \citenamefont {Díaz-Guilera}, \citenamefont {Kurths}, \citenamefont
  {Moreno},\ and\ \citenamefont {Zhou}}]{arenas2008}%
  \BibitemOpen
  \bibfield  {author} {\bibinfo {author} {\bibfnamefont {A.}~\bibnamefont
  {Arenas}}, \bibinfo {author} {\bibfnamefont {A.}~\bibnamefont
  {Díaz-Guilera}}, \bibinfo {author} {\bibfnamefont {J.}~\bibnamefont
  {Kurths}}, \bibinfo {author} {\bibfnamefont {Y.}~\bibnamefont {Moreno}}, \
  and\ \bibinfo {author} {\bibfnamefont {S.}~\bibnamefont {Zhou}},\ }\bibfield
  {title} {\enquote {\bibinfo {title} {{Synchronization in complex
  networks}},}\ }\href {\doibase 10.1016/j.physrep.2008.09.002} {\bibfield
  {journal} {\bibinfo  {journal} {Phys. Rep.}\ }\textbf {\bibinfo {volume}
  {469}},\ \bibinfo {pages} {93--153} (\bibinfo {year} {2008})}\BibitemShut
  {NoStop}%
\bibitem [{\citenamefont {Watanabe}\ and\ \citenamefont
  {Strogatz}(1994)}]{watanabe1994}%
  \BibitemOpen
  \bibfield  {author} {\bibinfo {author} {\bibfnamefont {S.}~\bibnamefont
  {Watanabe}}\ and\ \bibinfo {author} {\bibfnamefont {S.~H.}\ \bibnamefont
  {Strogatz}},\ }\bibfield  {title} {\enquote {\bibinfo {title} {{Constants of
  motion for superconducting Josephson arrays}},}\ }\href {\doibase
  10.1016/0167-2789(94)90196-1} {\bibfield  {journal} {\bibinfo  {journal}
  {Physica D}\ }\textbf {\bibinfo {volume} {74}},\ \bibinfo {pages} {197--253}
  (\bibinfo {year} {1994})}\BibitemShut {NoStop}%
\bibitem [{\citenamefont {Ott}\ and\ \citenamefont
  {Antonsen}(2008{\natexlab{a}})}]{ott2008}%
  \BibitemOpen
  \bibfield  {author} {\bibinfo {author} {\bibfnamefont {E.}~\bibnamefont
  {Ott}}\ and\ \bibinfo {author} {\bibfnamefont {T.~M.}\ \bibnamefont
  {Antonsen}},\ }\bibfield  {title} {\enquote {\bibinfo {title} {{Low
  dimensional behavior of large systems of globally coupled oscillators}},}\
  }\href {\doibase 10.1063/1.2930766} {\bibfield  {journal} {\bibinfo
  {journal} {Chaos}\ }\textbf {\bibinfo {volume} {18}},\ \bibinfo {pages}
  {037113} (\bibinfo {year} {2008}{\natexlab{a}})}\BibitemShut {NoStop}%
\bibitem [{\citenamefont {Marvel}, \citenamefont {Mirollo},\ and\ \citenamefont
  {Strogatz}(2009)}]{marvel2009}%
  \BibitemOpen
  \bibfield  {author} {\bibinfo {author} {\bibfnamefont {S.~A.}\ \bibnamefont
  {Marvel}}, \bibinfo {author} {\bibfnamefont {R.~E.}\ \bibnamefont {Mirollo}},
  \ and\ \bibinfo {author} {\bibfnamefont {S.~H.}\ \bibnamefont {Strogatz}},\
  }\bibfield  {title} {\enquote {\bibinfo {title} {{Identical phase oscillators
  with global sinusoidal coupling evolve by Möbius group action}},}\ }\href
  {\doibase 10.1063/1.3247089} {\bibfield  {journal} {\bibinfo  {journal}
  {Chaos}\ }\textbf {\bibinfo {volume} {19}},\ \bibinfo {pages} {043104}
  (\bibinfo {year} {2009})}\BibitemShut {NoStop}%
\bibitem [{\citenamefont {Chen}, \citenamefont {Engelbrecht},\ and\
  \citenamefont {Mirollo}(2017)}]{chen2017}%
  \BibitemOpen
  \bibfield  {author} {\bibinfo {author} {\bibfnamefont {B.}~\bibnamefont
  {Chen}}, \bibinfo {author} {\bibfnamefont {J.~R.}\ \bibnamefont
  {Engelbrecht}}, \ and\ \bibinfo {author} {\bibfnamefont {R.~E.}\ \bibnamefont
  {Mirollo}},\ }\bibfield  {title} {\enquote {\bibinfo {title} {{Hyperbolic
  geometry of Kuramoto oscillator networks}},}\ }\href {\doibase
  10.1088/1751-8121/aa7e39} {\bibfield  {journal} {\bibinfo  {journal} {J.
  Phys. A}\ }\textbf {\bibinfo {volume} {50}},\ \bibinfo {pages} {355101}
  (\bibinfo {year} {2017})}\BibitemShut {NoStop}%
\bibitem [{\citenamefont {Cestnik}\ and\ \citenamefont
  {Martens}(2024)}]{CestnikMartens2024}%
  \BibitemOpen
  \bibfield  {author} {\bibinfo {author} {\bibfnamefont {R.}~\bibnamefont
  {Cestnik}}\ and\ \bibinfo {author} {\bibfnamefont {E.}~\bibnamefont
  {Martens}},\ }\bibfield  {title} {\enquote {\bibinfo {title} {{Integrability
  of a Globally Coupled Complex Riccati Array: Quadratic Integrate-and-Fire
  Neurons, Phase Oscillators, and All in Between}},}\ }\href {\doibase
  10.1103/PhysRevLett.132.057201} {\bibfield  {journal} {\bibinfo  {journal}
  {Phys. Rev. Lett.}\ }\textbf {\bibinfo {volume} {132}},\ \bibinfo {pages}
  {057201} (\bibinfo {year} {2024})}\BibitemShut {NoStop}%
\bibitem [{\citenamefont {Kuramoto}(1975)}]{kuramoto1975}%
  \BibitemOpen
  \bibfield  {author} {\bibinfo {author} {\bibfnamefont {Y.}~\bibnamefont
  {Kuramoto}},\ }\href@noop {} {\emph {\bibinfo {title} {International
  Symposium on Mathematical Problems in Theoretical Physics}}}\ (\bibinfo
  {publisher} {Springer},\ \bibinfo {year} {1975})\ pp.\ \bibinfo {pages}
  {420--422}\BibitemShut {NoStop}%
\bibitem [{\citenamefont {Pikovsky}, \citenamefont {Rosenblum},\ and\
  \citenamefont {Kurths}(2001)}]{pikovsky2001}%
  \BibitemOpen
  \bibfield  {author} {\bibinfo {author} {\bibfnamefont {A.}~\bibnamefont
  {Pikovsky}}, \bibinfo {author} {\bibfnamefont {M.}~\bibnamefont {Rosenblum}},
  \ and\ \bibinfo {author} {\bibfnamefont {J.}~\bibnamefont {Kurths}},\
  }\href@noop {} {\emph {\bibinfo {title} {{Synchronization: A Universal
  Concept in Nonlinear Sciences}}}}\ (\bibinfo  {publisher} {Cambridge
  University Press},\ \bibinfo {year} {2001})\BibitemShut {NoStop}%
\bibitem [{\citenamefont {Vicsek}\ \emph {et~al.}(1995)\citenamefont {Vicsek},
  \citenamefont {Czirók}, \citenamefont {Ben-Jacob}, \citenamefont {Cohen},\
  and\ \citenamefont {Shochet}}]{vicsek1995}%
  \BibitemOpen
  \bibfield  {author} {\bibinfo {author} {\bibfnamefont {T.}~\bibnamefont
  {Vicsek}}, \bibinfo {author} {\bibfnamefont {A.}~\bibnamefont {Czirók}},
  \bibinfo {author} {\bibfnamefont {E.}~\bibnamefont {Ben-Jacob}}, \bibinfo
  {author} {\bibfnamefont {I.}~\bibnamefont {Cohen}}, \ and\ \bibinfo {author}
  {\bibfnamefont {O.}~\bibnamefont {Shochet}},\ }\bibfield  {title} {\enquote
  {\bibinfo {title} {{Novel type of phase transition in a system of self-driven
  particles}},}\ }\href {\doibase 10.1103/PhysRevLett.75.1226} {\bibfield
  {journal} {\bibinfo  {journal} {Phys. Rev. Lett.}\ }\textbf {\bibinfo
  {volume} {75}},\ \bibinfo {pages} {1226} (\bibinfo {year}
  {1995})}\BibitemShut {NoStop}%
\bibitem [{\citenamefont {Rohden}\ \emph {et~al.}(2012)\citenamefont {Rohden},
  \citenamefont {Sorge}, \citenamefont {Timme},\ and\ \citenamefont
  {Witthaut}}]{rohden2012}%
  \BibitemOpen
  \bibfield  {author} {\bibinfo {author} {\bibfnamefont {M.}~\bibnamefont
  {Rohden}}, \bibinfo {author} {\bibfnamefont {A.}~\bibnamefont {Sorge}},
  \bibinfo {author} {\bibfnamefont {M.}~\bibnamefont {Timme}}, \ and\ \bibinfo
  {author} {\bibfnamefont {D.}~\bibnamefont {Witthaut}},\ }\bibfield  {title}
  {\enquote {\bibinfo {title} {{Self-Organized Synchronization in Decentralized
  Power Grids}},}\ }\href {\doibase 10.1103/PhysRevLett.109.064101} {\bibfield
  {journal} {\bibinfo  {journal} {Phys. Rev. Lett.}\ }\textbf {\bibinfo
  {volume} {109}},\ \bibinfo {pages} {064101} (\bibinfo {year}
  {2012})}\BibitemShut {NoStop}%
\bibitem [{\citenamefont {Bick}\ \emph {et~al.}(2020)\citenamefont {Bick},
  \citenamefont {Goodfellow}, \citenamefont {Laing},\ and\ \citenamefont
  {Martens}}]{Bick2018c}%
  \BibitemOpen
  \bibfield  {author} {\bibinfo {author} {\bibfnamefont {C.}~\bibnamefont
  {Bick}}, \bibinfo {author} {\bibfnamefont {M.}~\bibnamefont {Goodfellow}},
  \bibinfo {author} {\bibfnamefont {C.~R.}\ \bibnamefont {Laing}}, \ and\
  \bibinfo {author} {\bibfnamefont {E.~A.}\ \bibnamefont {Martens}},\
  }\bibfield  {title} {\enquote {\bibinfo {title} {{Understanding the dynamics
  of biological and neural oscillator networks through exact mean-field
  reductions: a review}},}\ }\href {\doibase 10.1186/s13408-020-00086-9}
  {\bibfield  {journal} {\bibinfo  {journal} {J. Math. Neurosci.}\ }\textbf
  {\bibinfo {volume} {10}},\ \bibinfo {pages} {9} (\bibinfo {year}
  {2020})}\BibitemShut {NoStop}%
\bibitem [{\citenamefont {Ince}(1956)}]{inceordinary}%
  \BibitemOpen
  \bibfield  {author} {\bibinfo {author} {\bibfnamefont {E.~L.}\ \bibnamefont
  {Ince}},\ }\href@noop {} {\emph {\bibinfo {title} {{Ordinary Differential
  Equations}}}}\ (\bibinfo  {publisher} {Dover},\ \bibinfo {year} {1956})\ pp.\
  \bibinfo {pages} {23--35 and 295}\BibitemShut {NoStop}%
\bibitem [{\citenamefont {Lohe}(2018)}]{Lohe18}%
  \BibitemOpen
  \bibfield  {author} {\bibinfo {author} {\bibfnamefont {M.~A.}\ \bibnamefont
  {Lohe}},\ }\bibfield  {title} {\enquote {\bibinfo {title}
  {{Higher-dimensional generalizations of the Watanabe–Strogatz transform for
  vector models of synchronization}},}\ }\href {\doibase
  10.1088/1751-8121/aac030} {\bibfield  {journal} {\bibinfo  {journal} {J.
  Phys. A}\ }\textbf {\bibinfo {volume} {51}},\ \bibinfo {pages} {225101}
  (\bibinfo {year} {2018})}\BibitemShut {NoStop}%
\bibitem [{\citenamefont {Lohe}(2019)}]{Lohe19}%
  \BibitemOpen
  \bibfield  {author} {\bibinfo {author} {\bibfnamefont {M.~A.}\ \bibnamefont
  {Lohe}},\ }\bibfield  {title} {\enquote {\bibinfo {title} {{Systems of matrix
  Riccati equations, linear fractional transformations, partial integrability
  and synchronization}},}\ }\href {\doibase 10.1063/1.5085248} {\bibfield
  {journal} {\bibinfo  {journal} {J. Math. Phys.}\ }\textbf {\bibinfo {volume}
  {60}} (\bibinfo {year} {2019}),\ 10.1063/1.5085248}\BibitemShut {NoStop}%
\bibitem [{\citenamefont {Mirollo}\ and\ \citenamefont
  {Strogatz}(1991)}]{mirollo1991stability}%
  \BibitemOpen
  \bibfield  {author} {\bibinfo {author} {\bibfnamefont {R.~E.}\ \bibnamefont
  {Mirollo}}\ and\ \bibinfo {author} {\bibfnamefont {S.~H.}\ \bibnamefont
  {Strogatz}},\ }\bibfield  {title} {\enquote {\bibinfo {title} {{Stability of
  incoherence in a population of coupled oscillators}},}\ }\href {\doibase
  10.1007/BF01029202} {\bibfield  {journal} {\bibinfo  {journal} {J. Stat.
  Phys.}\ }\textbf {\bibinfo {volume} {63}},\ \bibinfo {pages} {613--635}
  (\bibinfo {year} {1991})}\BibitemShut {NoStop}%
\bibitem [{\citenamefont {Paz\'o}\ and\ \citenamefont
  {Cestnik}(2025)}]{PazoCestnik25}%
  \BibitemOpen
  \bibfield  {author} {\bibinfo {author} {\bibfnamefont {D.}~\bibnamefont
  {Paz\'o}}\ and\ \bibinfo {author} {\bibfnamefont {R.}~\bibnamefont
  {Cestnik}},\ }\bibfield  {title} {\enquote {\bibinfo {title}
  {{Low-dimensional dynamics of globally coupled complex Riccati equations:
  Exact firing-rate equations for spiking neurons with clustered
  substructure}},}\ }\href {\doibase 10.1103/PhysRevE.111.L052201} {\bibfield
  {journal} {\bibinfo  {journal} {Phys. Rev. E}\ }\textbf {\bibinfo {volume}
  {111}},\ \bibinfo {pages} {L052201} (\bibinfo {year} {2025})}\BibitemShut
  {NoStop}%
\bibitem [{\citenamefont {Ott}\ and\ \citenamefont
  {Antonsen}(2008{\natexlab{b}})}]{OA08}%
  \BibitemOpen
  \bibfield  {author} {\bibinfo {author} {\bibfnamefont {E.}~\bibnamefont
  {Ott}}\ and\ \bibinfo {author} {\bibfnamefont {T.~M.}\ \bibnamefont
  {Antonsen}},\ }\bibfield  {title} {\enquote {\bibinfo {title} {{Low
  dimensional behavior of large systems of globally coupled oscillators}},}\
  }\href {\doibase 10.1063/1.2930766} {\bibfield  {journal} {\bibinfo
  {journal} {Chaos}\ }\textbf {\bibinfo {volume} {18}},\ \bibinfo {eid}
  {037113} (\bibinfo {year} {2008}{\natexlab{b}})}\BibitemShut {NoStop}%
\bibitem [{\citenamefont {Montbri\'o}, \citenamefont {Paz\'o},\ and\
  \citenamefont {Roxin}(2015)}]{LA15}%
  \BibitemOpen
  \bibfield  {author} {\bibinfo {author} {\bibfnamefont {E.}~\bibnamefont
  {Montbri\'o}}, \bibinfo {author} {\bibfnamefont {D.}~\bibnamefont {Paz\'o}},
  \ and\ \bibinfo {author} {\bibfnamefont {A.}~\bibnamefont {Roxin}},\
  }\bibfield  {title} {\enquote {\bibinfo {title} {{Macroscopic Description for
  Networks of Spiking Neurons}},}\ }\href {\doibase 10.1103/PhysRevX.5.021028}
  {\bibfield  {journal} {\bibinfo  {journal} {Phys. Rev. X}\ }\textbf {\bibinfo
  {volume} {5}},\ \bibinfo {pages} {021028} (\bibinfo {year}
  {2015})}\BibitemShut {NoStop}%
\bibitem [{\citenamefont {Laing}(2015)}]{Laing15}%
  \BibitemOpen
  \bibfield  {author} {\bibinfo {author} {\bibfnamefont {C.~R.}\ \bibnamefont
  {Laing}},\ }\bibfield  {title} {\enquote {\bibinfo {title} {{Exact Neural
  Fields Incorporating Gap Junctions}},}\ }\href {\doibase 10.1137/15M1011287}
  {\bibfield  {journal} {\bibinfo  {journal} {SIAM J. Appl. Dyn. Syst.}\
  }\textbf {\bibinfo {volume} {14}},\ \bibinfo {pages} {1899--1929} (\bibinfo
  {year} {2015})}\BibitemShut {NoStop}%
\bibitem [{\citenamefont {Cestnik}\ and\ \citenamefont
  {Pikovsky}(2022{\natexlab{a}})}]{CestnikPikovsky2022}%
  \BibitemOpen
  \bibfield  {author} {\bibinfo {author} {\bibfnamefont {R.}~\bibnamefont
  {Cestnik}}\ and\ \bibinfo {author} {\bibfnamefont {A.}~\bibnamefont
  {Pikovsky}},\ }\bibfield  {title} {\enquote {\bibinfo {title} {{Hierarchy of
  Exact Low-Dimensional Reductions for Populations of Coupled Oscillators}},}\
  }\href {\doibase 10.1103/PhysRevLett.128.054101} {\bibfield  {journal}
  {\bibinfo  {journal} {Phys. Rev. Lett.}\ }\textbf {\bibinfo {volume} {128}},\
  \bibinfo {pages} {054101} (\bibinfo {year} {2022}{\natexlab{a}})}\BibitemShut
  {NoStop}%
\bibitem [{\citenamefont {Cestnik}\ and\ \citenamefont
  {Pikovsky}(2022{\natexlab{b}})}]{CestnikPikovsky2022b}%
  \BibitemOpen
  \bibfield  {author} {\bibinfo {author} {\bibfnamefont {R.}~\bibnamefont
  {Cestnik}}\ and\ \bibinfo {author} {\bibfnamefont {A.}~\bibnamefont
  {Pikovsky}},\ }\bibfield  {title} {\enquote {\bibinfo {title} {{Exact
  finite-dimensional reduction for a population of noisy oscillators and its
  link to Ott-Antonsen and Watanabe-Strogatz theories}},}\ }\href {\doibase
  10.1063/5.0106171} {\bibfield  {journal} {\bibinfo  {journal} {Chaos}\
  }\textbf {\bibinfo {volume} {32}},\ \bibinfo {pages} {113126} (\bibinfo
  {year} {2022}{\natexlab{b}})}\BibitemShut {NoStop}%
\bibitem [{\citenamefont {Pietras}, \citenamefont {Cestnik},\ and\
  \citenamefont {Pikovsky}(2023)}]{PietrasPikovsky2023}%
  \BibitemOpen
  \bibfield  {author} {\bibinfo {author} {\bibfnamefont {B.}~\bibnamefont
  {Pietras}}, \bibinfo {author} {\bibfnamefont {R.}~\bibnamefont {Cestnik}}, \
  and\ \bibinfo {author} {\bibfnamefont {A.}~\bibnamefont {Pikovsky}},\
  }\bibfield  {title} {\enquote {\bibinfo {title} {{Exact finite-dimensional
  description for networks of globally coupled spiking neurons}},}\ }\href
  {\doibase 10.1103/PhysRevE.107.024315} {\bibfield  {journal} {\bibinfo
  {journal} {Phys. Rev. E}\ }\textbf {\bibinfo {volume} {107}},\ \bibinfo
  {pages} {024315} (\bibinfo {year} {2023})}\BibitemShut {NoStop}%
\bibitem [{\citenamefont {T\"onjes}\ and\ \citenamefont
  {Pikovsky}(2020)}]{ToenjesPikovksy2020}%
  \BibitemOpen
  \bibfield  {author} {\bibinfo {author} {\bibfnamefont {R.}~\bibnamefont
  {T\"onjes}}\ and\ \bibinfo {author} {\bibfnamefont {A.}~\bibnamefont
  {Pikovsky}},\ }\bibfield  {title} {\enquote {\bibinfo {title}
  {{Low-dimensional description for ensembles of identical phase oscillators
  subject to Cauchy noise}},}\ }\href {\doibase 10.1103/PhysRevE.102.052315}
  {\bibfield  {journal} {\bibinfo  {journal} {Phys. Rev. E}\ }\textbf {\bibinfo
  {volume} {102}},\ \bibinfo {pages} {052315} (\bibinfo {year}
  {2020})}\BibitemShut {NoStop}%
\bibitem [{\citenamefont {Clusella}\ and\ \citenamefont
  {Montbri\'o}(2024)}]{ClusellaMontbrio2024}%
  \BibitemOpen
  \bibfield  {author} {\bibinfo {author} {\bibfnamefont {P.}~\bibnamefont
  {Clusella}}\ and\ \bibinfo {author} {\bibfnamefont {E.}~\bibnamefont
  {Montbri\'o}},\ }\bibfield  {title} {\enquote {\bibinfo {title} {{Exact
  low-dimensional description for fast neural oscillations with low firing
  rates}},}\ }\href {\doibase 10.1103/PhysRevE.109.014229} {\bibfield
  {journal} {\bibinfo  {journal} {Phys. Rev. E}\ }\textbf {\bibinfo {volume}
  {109}},\ \bibinfo {pages} {014229} (\bibinfo {year} {2024})}\BibitemShut
  {NoStop}%
\end{thebibliography}

%merlin.mbs aipnum4-1.bst 2010-07-25 4.21a (PWD, AO, DPC) hacked
%Control: key (0)
%Control: author (8) initials jnrlst
%Control: editor formatted (1) identically to author
%Control: production of article title (0) allowed
%Control: page (1) range
%Control: year (1) truncated
%Control: production of eprint (0) enabled
%

\end{document}